\input{aipcheck}

\documentclass[
    ,final            % use final for the camera ready runs
  ,numberedheadings % uncomment this option for numbered sections
  ]
  {aipproc}

\layoutstyle{6x9}
\usepackage{epstopdf}
\usepackage{graphicx}
\begin{document}

\title{Nuclear charge radii and electric monopole transitions
in the interacting boson model}


\author{P.~Van~Isacker}
{address={Grand Acc\'el\'erateur National d'Ions Lourds, CEA/DSM--CNRS/IN2P3\\
Boulevard Henri Becquerel, BP~55027, F-14076 Caen Cedex 5, France}}

\begin{abstract}
The interacting boson model (IBM) of Arima and Iachello is applied
to calculate nuclear charge radii and electric monopole transitions
of even-even nuclei in the rare-earth region.
Consistent operators are used for the two observables.
A relation between summed M1 strength and $\rho({\rm E0})$ values is pointed out.
\end{abstract}

\maketitle

\section{Introduction}
\label{s_intro}
According to the geometric model of Bohr and Mottelson~\cite{Bohr75},
a nucleus with an elipsoidal equilibrium shape
may undergo oscillations of two different types, $\beta$ and $\gamma$.
While $\gamma$ vibrations
are an acknowledged feature of deformed nuclei,
such is not the case for the $\beta$ kind.
A careful analysis of the observed properties of excited $0^+$ states
seems to indicate that very few indeed
satisfy all criteria proper to a $\beta$-vibrational state~\cite{Garrett01}.
In particular, very few decay to the ground state
by way of an E0 transition of sizable strength~\cite{Wood99},
as should be for a $\beta$ vibration~\cite{Reiner61}.
It is therefore not surprising
that alternative interpretations of excited $0^+$ states in deformed nuclei,
either as pairing isomers~(see, {\it e.g.}, Refs.~\cite{Kulp03,Kulp05})
or through shape coexistence and configuration mixing (see, {\it e.g.}, Ref.~\cite{Kulp08})
have gained advocates over recent years.

In this contribution it is shown
that a collective interpretation of the ground and excited $0^+$ states
with the interacting boson model (IBM)
of Arima and Iachello~\cite{Arima76}
can account for the data on charge radii and E0 transitions,
as observed in the rare-earth region.
Full details of this work can be found in Refs.~\cite{Zerguine08,Zerguine12}.
In addition, a correlation with summed M1 strength is pointed out.

\section{Charge radii and electric monopole transitions}
\label{s_r2e0}
The probability for an E0 transition to occur
between an initial state $|i\rangle$ and a final state $|f\rangle$
can be written as the product of an electronic factor $\Omega$
and the square of a nuclear factor $\rho({\rm E0})$,
the latter being equal to
\begin{equation}
\rho({\rm E0};i\rightarrow f)=
\frac{|\langle f|\hat T({\rm E0})|i\rangle|}{eR^2},
\label{e_rho}
\end{equation}
with $R=r_0A^{1/3}$.
In first approximation the E0 matrix element equals
\begin{equation}
\langle f|\hat T_{\rm p}({\rm E0})|i\rangle=
\langle f|e\sum_{k\in{\rm p}}r^2_k|i\rangle.
\label{e_e0a}
\end{equation}
where the sum is over all protons in the nucleus~\cite{Church56}.
In the nuclei considered here not all protons can be treated explicitly
and recourse should be taken to effective charges
$e_{\rm n}$ and $e_{\rm p}$ for the neutrons (n) and protons (p),
leading to the generalized expression~\cite{Kantele84}
\begin{equation}
\langle f|\hat T({\rm E0})|i\rangle=
\langle f|\sum_{k=1}^Ae_kr^2_k|i\rangle=
\langle f|\left(e_{\rm n}\sum_{k\in{\rm n}}r^2_k+e_{\rm p}\sum_{k\in{\rm p}}r^2_k\right)|i\rangle.
\label{e_e0}
\end{equation}
On the other hand,
the mean-square charge radius of a state $|s\rangle$
is given by
\begin{equation}
\langle s|\hat T_{\rm p}(r^2)|s\rangle=
\frac{1}{Z}\langle s|\sum_{k\in{\rm p}}r^2_k|s\rangle.
\label{e_r2a}
\end{equation}
This is an appropriate expression
insofar that a realistic $A$-body wave function
is used for the state $|s\rangle$.
This is often impossible
and an effective charge radius operator $\hat T(r^2)$ should then be taken. 
The generalization of the  expression~(\ref{e_r2a}),
similar to the one carried out for the E0 operator,
can therefore be written as
\begin{equation}
\langle r^2 \rangle_s\equiv
\langle s|\hat T(r^2)|s\rangle=
\frac{1}{e_{\rm n}N+e_{\rm p}Z}
\langle s|\sum_{k=1}^Ae_kr^2_k|s\rangle.
\label{e_r2}
\end{equation}
The basic hypothesis of the present study
is to assume that {\em the effective nucleon charges
in the charge radius and E0 transition operators are the same}.
If this is so, comparison of Eqs.~(\ref{e_e0}) and~(\ref{e_r2})
leads to the operator relation
\begin{equation}
\hat T({\rm E0})=(e_{\rm n}N+e_{\rm p}Z)\hat T(r^2).
\label{e_e0r2}
\end{equation}
This is a general relation between the effective operators
to be used for the calculation of charge radii and E0 transitions.
Equation~(\ref{e_e0r2}) can, in principle, be tested in the framework of any model
and here the implied correlation is tested
in the framework of the IBM~\cite{Arima76}.
This requires that all states involved
[{\it i.e.}, $|i\rangle$ and $|f\rangle$ in Eq.~(\ref{e_rho}),
and $|s\rangle$  in Eq.~(\ref{e_r2})]
are collective in character
and can be described by the IBM.

In the \mbox{IBM-1} the charge radius operator
is taken as the most general scalar expression,
linear in the generators of U(6)~\cite{Iachello87},
\begin{equation}
\hat T(r^2)=
\langle r^2\rangle_{\rm c}+
\alpha N_{\rm b}+
\eta\frac{\hat n_d}{N_{\rm b}},
\label{e_r2sd}
\end{equation}
where $N_{\rm b}$ is the total boson number,
$\hat n_d$ is the $d$-boson number operator,
and $\alpha$ and $\eta$ are parameters with units of length$^2$.
The first term in Eq.~(\ref{e_r2sd}), $\langle r^2\rangle_{\rm c}$,
is the square of the charge radius of the core nucleus.
The second term accounts for the (locally linear) increase
in the charge radius due to the addition of two nucleons
({\it i.e.}, neutrons since isotope shifts are considered in this study).
The third term in Eq.~(\ref{e_r2sd})
stands for the contribution to the charge radius due to deformation.
It is identical to the one given in Ref.~\cite{Iachello87}
but for the factor $1/N_{\rm b}$.
This factor is included here
because it is the {\em fraction} $\langle\hat n_d\rangle/N_{\rm b}$
which is a measure of the quadrupole deformation
($\beta_2^2$ in the geometric collective model)
rather than the matrix element $\langle\hat n_d\rangle$ itself.

Two quantities can be derived from charge radii,
namely isotope and isomer shifts.
The former measure the difference in charge radius of neighboring isotopes.
For the difference between even-even isotopes
one finds from Eq.~(\ref{e_r2sd}) 
\begin{equation}
\Delta \langle r^2\rangle\equiv
\langle r^2\rangle_{0_1^+}^{(A+2)}-\langle r^2\rangle_{0_1^+}^{(A)}=
|\alpha|+\eta
\left(\langle\frac{\hat n_d}{N_{\rm b}}\rangle_{0_1^+}^{(A+2)}-
\langle\frac{\hat n_d}{N_{\rm b}}\rangle_{0_1^+}^{(A)}\right).
\label{e_ips}
\end{equation}
Isomer shifts are a measure of the difference in charge radius
between an excited (here the $2_1^+$) state
and the ground state, and are given by
\begin{equation}
\delta\langle r^2\rangle\equiv
\langle r^2\rangle^{(A)}_{2_1^+}-\langle r^2\rangle^{(A)}_{0_1^+}=
\frac{\eta}{N_{\rm b}}
\left(\langle\hat n_d\rangle_{2_1^+}^{(A)}-
\langle\hat n_d\rangle_{0_1^+}^{(A)}\right).
\label{e_ims}
\end{equation}

Once the form of the charge radius operator is determined,
the E0 transition operator follows from Eq.~(\ref{e_e0r2}).
In the \mbox{IBM-1} the E0 transition operator is therefore
\begin{equation}
\hat T({\rm E0})=
(e_{\rm n}N+e_{\rm p}Z)
\eta\frac{\hat n_d}{N_{\rm b}}.
\label{e_e0sd}
\end{equation}
Since for E0 transitions the initial and final states are different,
neither the constant $\langle r^2\rangle_{\rm c}$
nor $\alpha N_{\rm b}$ in Eq.~(\ref{e_r2sd}) contribute to the transition,
and its $\rho({\rm E0})$ value equals
\begin{equation}
\rho({\rm E0};i\rightarrow f)=
\frac{e_{\rm n}N+e_{\rm p}Z}{eR^2}
\frac{\eta}{N_{\rm b}}
|\langle f|\hat n_d|i\rangle|,
\label{e_rhosd}
\end{equation}

\section{Application to the rare-earth region}
\label{s_application}
To test the relation between charge radii and E0 transitions,
proposed in the previous section, 
a systematic study of all even-even isotopic chains
from Ce ($Z=58$) to W ($Z=74$) is carried out.
This analysis requires the knowledge of structural information
concerning the ground and excited states
which is obtained by adjusting an \mbox{IBM-1} Hamiltonian
to observed spectra in the rare-earth region.
The details of the energy calculation
can be found in Ref.~\cite{Zerguine12}.
The procedure is closely related to the one followed
by Garc\'\i a-Ramos {\it et al.}~\cite{Garcia03}
and yields a root-mean square deviation for an entire isotopic chain
which is typically of the order of 100~keV.

\begin{figure}
\includegraphics[width=14cm]{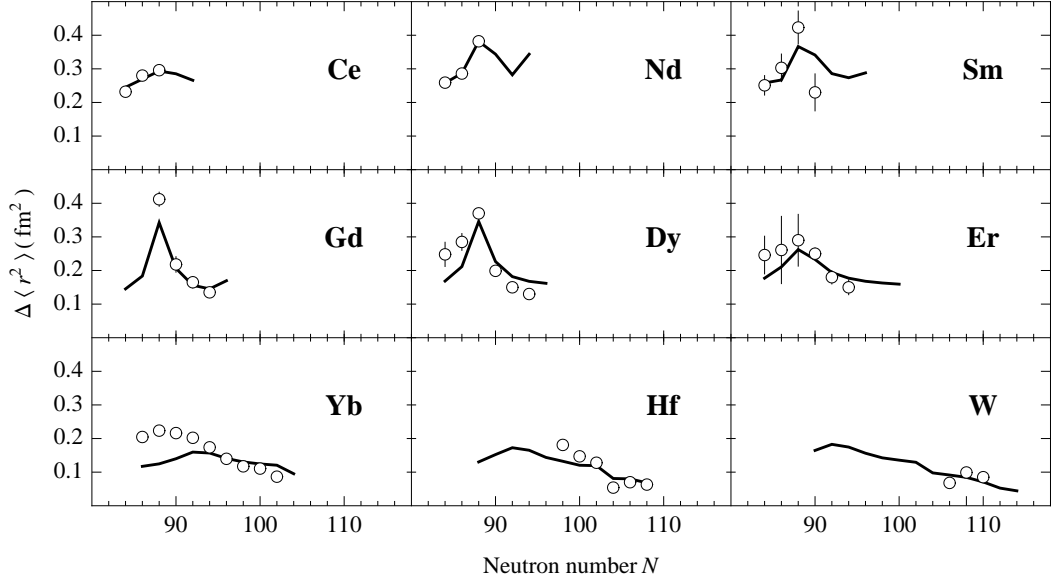}
\caption{
Isotope shifts $\Delta\langle r^2\rangle$ in the rare-earth region.
The points are the experimental values
and the lines are calculated with Eq.~(\ref{e_ips}).
References to experiments are given in Ref.~\cite{Zerguine12}.}
\label{f_tshift} 
\end{figure}
Isotope shifts $\Delta\langle r^2\rangle$, according to Eq.~(\ref{e_ips}),
depend on the parameters $|\alpha|$ and $\eta$
in the $\mbox{IBM-1}$ operator~(\ref{e_r2sd}). 
The parameter $|\alpha|$ is adjusted
for each isotope series separately,
while $\eta$ is kept constant for all isotopes,
$\eta=0.50$~fm$^2$.
The resulting isotope shifts are shown in Fig.~\ref{f_tshift}.
The peaks in the isotope shifts are well reproduced 
in all isotopic chains with the exception of Yb.
The largest peaks occur for $^{152-150}$Sm, $^{154-152}$Gd, and $^{156-154}$Dy,
that is, for the difference in radii between $N=90$ and $N=88$ isotopes.
The peak is smaller below $Z=62$ for Ce and Nd,
and fades away above $Z=66$ for Er, Yb, Hf, and W.
The calculated isotope shifts broadly agree with these observed features
but there are differences though.
Notably, the calculated peak in the Sm isotopes
is much broader than the observed one,
indicating that the spherical-to-deformed transition
occurs faster in reality than it does in the \mbox{IBM-1} calculation.

\begin{figure}
\includegraphics[width=10cm]{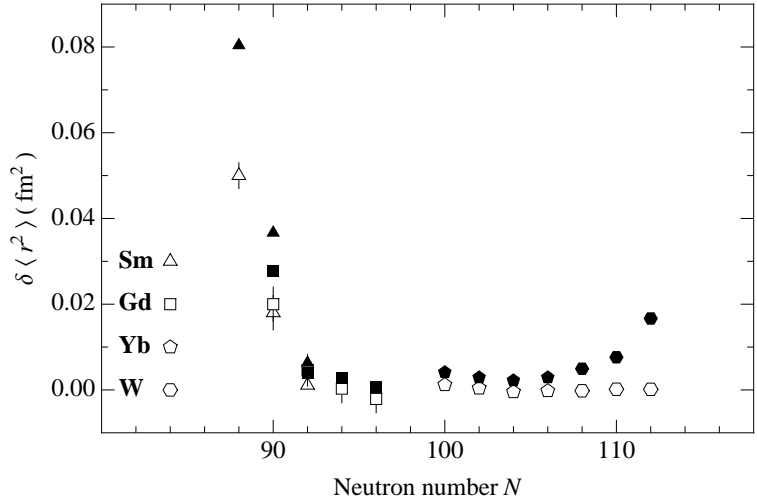}
\caption{
Isomer shifts $\delta\langle r^2\rangle$ in the rare-earth region.
The open symbols are the experimental values
and the full symbols are calculated with Eq.~(\ref{e_ims}).
References to experiments are given in Ref.~\cite{Zerguine12}.}
\label{f_mshift} 
\end{figure}
A further test of the calculated charge radii
is obtained from isomer shifts $\delta\langle r^2\rangle$,
depending only on $\eta$ [see Eq.~(\ref{e_ims})].
The isomer shifts known experimentally are shown in Fig.~\ref{f_mshift}.
The data are more than 30 years old and often discrepant,
in particular in the Yb and W isotopes.
Nevertheless, a clear conclusion can be drawn
from the isomer shifts measured in the Sm and Gd isotopes:
they are an order of magnitude smaller in the deformed
than they are in the spherical region.

\begin{figure}
\includegraphics[width=10cm]{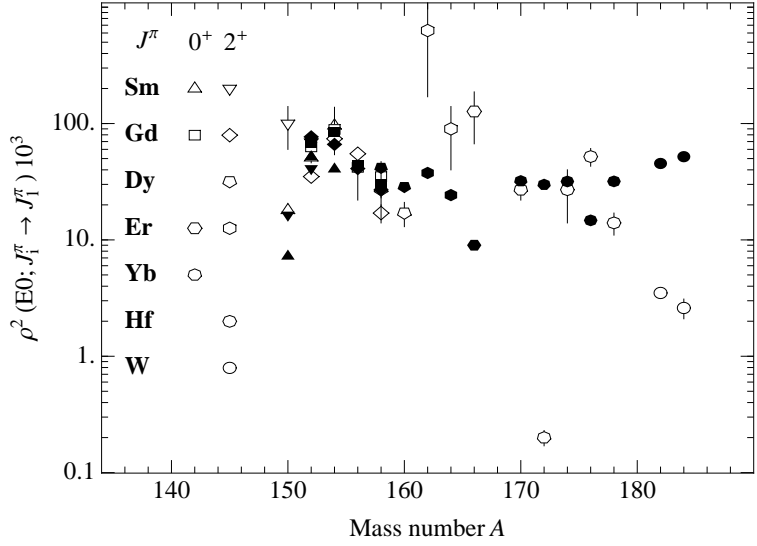}
\caption{
Electric monopole $\rho^2({\rm E0};J^\pi_i\rightarrow J^\pi_1)$ values
(times $10^3$) in the rare-earth region
for $J^\pi=0^+$ and $2^+$.
The open symbols are the experimental values
and the full symbols are calculated with Eq.~(\ref{e_rhosd}).
References to experiments are given in Ref.~\cite{Zerguine12}.}
\label{f_rho2} 
\end{figure}
The question is now whether the value $\eta=0.50$~fm$^2$,
obtained from isotope and isomer shifts,
reproduces the E0 transitions observed in the rare-earth nuclei.
In Fig.~\ref{f_rho2} the available E0 data
for $0^+_i\rightarrow0^+_1$ and $2^+_i\rightarrow2^+_1$ transitions in the rare-earth region
are compared with the results of the \mbox{IBM-1} calculation.
It is not the intention to give a detailed comparison for all transitions
(which can be found in Ref.~\cite{Zerguine12})
but the overall impression is that
the present approach succeeds in reproducing
the correct order of magnitude for $\rho^2({\rm E}0)$.
However, some discrepancies can be observed in heavier nuclei
and especially concern $^{172}$Yb and $^{182-184}$W.
A possible explanation is
that the $\rho^2({\rm E}0)$ measured for these nuclei
is not associated with collective states.
This seems to be the case in $^{172}$Yb
where several $\rho^2({\rm E}0)$ have been measured,
none of which is large.
Only in the W isotopes does it seem that the observed E0 strength
is an order of magnitude smaller than the calculated value.
It is known that these nuclei
are in a region of hexadecapole deformation~\cite{Lee74}
and this may offer a qualitative explanation
of the suppression of the E0 strength~\cite{Zerguine12}.

\section{Relation to summed M1 strength}
\label{s_m1}
After the presentation of this work during the conference,
J.N.~Ginocchio asked a question
concerning a possible correlation
with summed M1 strength to the scissors mode.
The argument goes as follows.
It is known from the work of Ginocchio~\cite{Ginocchio91}
that the summed M1 strength from the ground state to the scissors mode
(for a review on the latter, see Ref.~\cite{Heyde10})
is related to the ground-state matrix element
of the $d$-boson number operator $\hat n_d$,
\begin{equation}
\sum_f
B({\rm M1};0^+_1\rightarrow1^+_f)=
\frac{9}{8\pi}(g_\nu-g_\pi)^2
\frac{n\;z}{N_{\rm b}(N_{\rm b}-1)}
\langle0^+_1|\hat n_d|0^+_1\rangle,
\label{e_bm1}
\end{equation}
where $n$ ($z$) is the number of valence neutron (proton) particles or holes,
whichever is smaller,
and $g_\nu$ and $g_\pi$ are the $g$-factors
of the neutron and proton bosons.
Note that $n+z=2N_{\rm b}$.
The relation~(\ref{e_bm1}) was used in Ref.~\cite{Heyde93}
to establish a connection between the summed M1 strength
and the charge radii of the Nd, Sm, and Dy nuclei.
It is therefore natural to ask
whether in turn a connection exists
between the summed M1 strength and $\rho({\rm E0})$ values.

Such a connection can indeed be established.
The E0 operator is directly proportional to $\hat n_d$,
unlike the charge-radius operator
which involves additional terms [see Eq.~(\ref{e_r2sd})]
which complicate the relation
between the summed M1 strength and charge radii. 
On the other hand, the matrix element of $\hat n_d$
appearing in the sum rule~(\ref{e_bm1}) is diagonal
while a $\rho({\rm E0})$ value
involves a non-diagonal matrix element of $\hat n_d$.
A relation between the two matrix elements
can nevertheless be obtained in the SU(3) limit
where they are~\cite{Subber88}
\begin{equation}
\langle0^+_1|\hat n_d|0^+_1\rangle=
\frac{4N_{\rm b}(N_{\rm b}-1)}{3(2N_{\rm b}-1)},
\qquad
|\langle0^+_\beta|\hat n_d|0^+_1\rangle|=
\left[\frac{8(N_{\rm b}-1)^2N_{\rm b}(2N_{\rm b}+1)}{9(2N_{\rm b}-3)(2N_{\rm b}-1)^2}\right]^{1/2}.
\label{e_su3mes}
\end{equation}
From these expressions the ratio of matrix elements can be derived,
resulting in the following relation,
valid in the large-$N_{\rm b}$ limit:
\begin{equation}
B({\rm M}1;0^+_1\rightarrow1^+_1)\approx
\frac{9}{4\pi}(g_\nu-g_\pi)^2\frac{r_0^2}{\eta}
g(N,Z,n,z)\rho({\rm E}0;0^+_\beta\rightarrow0^+_1),
\label{e_rhobm1}
\end{equation}
where $g(N,Z,n,z)$ is the function
\begin{equation}
g(N,Z,n,z)=
\frac{e(N+Z)^{2/3}}{e_{\rm n}N+e_{\rm p}Z}
\frac{n\;z}{\sqrt{n+z}}.
\label{e_gfun}
\end{equation}
Note that in the SU(3) limit only one $1^+$ state is excited
and only the $\beta$-vibrational state decays by E0 to the ground state,
as is indicated in Eq.~(\ref{e_rhobm1}).

\begin{figure}
\includegraphics[width=10cm]{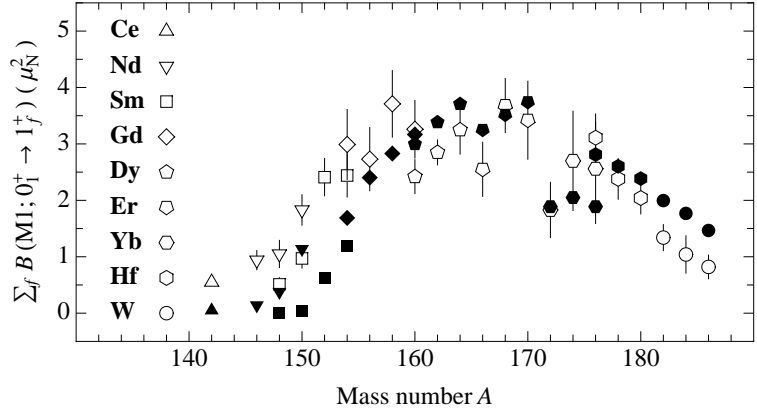}
\caption{
Summed M1 strength
$\sum_f B({\rm M1};0^+_1\rightarrow1^+_f)$ in the rare-earth region.
The open symbols are the experimental values~\cite{Enders05}
and the full symbols are calculated with Eq.~(\ref{e_bm1}).}
\label{f_bm1} 
\end{figure}
It should be emphasized that
the relation~(\ref{e_rhobm1}) is valid only in the SU(3) limit
which might jeopardize its use in transitional nuclei.
One may nevertheless attempt to apply it to the entire rare-earth region.
The ratio $r_0^2/\eta=3.08$
and the effective charges $e_{\rm n}=0.5e$ and $e_{\rm p}=e$
have been determined from a fit to radii~\cite{Zerguine12}.
The only remaining constant in Eq.~(\ref{e_rhobm1}), $g_\nu-g_\pi$,
can be obtained by adjusting the expression~(\ref{e_bm1})
to the observed summed M1 strength in rare-earth nuclei~\cite{Enders05},
as shown in Fig.~\ref{f_bm1}.
This, incidentally, constitutes an additional test
of the matrix elements $\langle0^+_1|\hat n_d|0^+_1\rangle$
as they are obtained in the \mbox{IBM-1} fits
to the different isotope series.
An overall satisfactory agreement is obtained
with the value $|g_\nu-g_\pi|=0.83$~$\mu_{\rm N}$
but it is seen that the calculated M1 strength in the Sm isotopes
is well below its observed value,
again indicating that the calculated transition
is too slow for this isotope series.

\begin{figure}
\includegraphics[width=10cm]{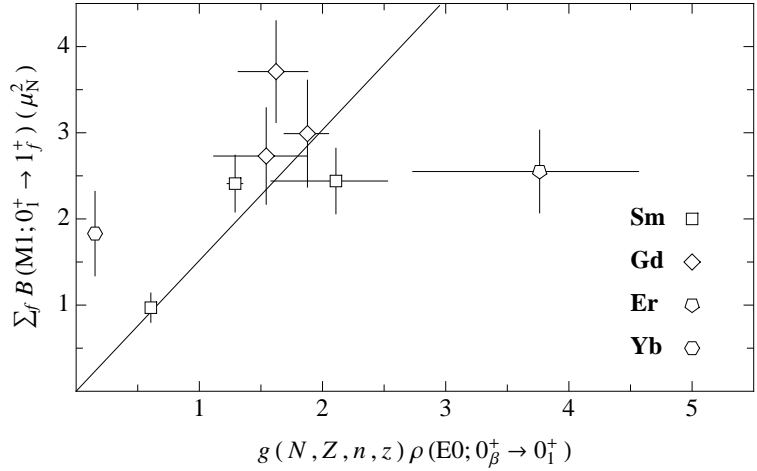}
\caption{
Correlation between the $\rho({\rm E0};0^+_\beta\rightarrow 0^+_1)$ value
and the summed M1 strength $\sum_f B({\rm M1};0^+_1\rightarrow1^+_f)$
for the eight nuclei in the rare-earth region
where both properties are known.
The function $g(N,Z,n,z)$ is defined in Eq.~(\ref{e_gfun}).
The slope of the line is $9(g_\nu-g_\pi)^2r_0^2/(4\pi\eta)$~$\mu_{\rm N}^2$.}
\label{f_rhobm1} 
\end{figure}
The correlation~(\ref{e_rhobm1}) can now be tested
and is shown in Fig.~\ref{f_rhobm1}
for the eight nuclei in the rare-earth region
where both E0 and M1 properties are known
($^{150,152,154}$Sm, $^{154,156,158}$Gd, $^{166}$Er, and $^{172}$Yb).
The $^{172}$Yb point is conspicuously off the line,
reinforcing the earlier remark about E0 strength in this nucleus.
The $^{166}$Er point follows from a recent experiment~\cite{Wimmer09} 
where the {\em fourth} $J^\pi=0^+$ level at 1934~keV
has been identified as the band head of the $\beta$-vibrational band
with a sizable E0 matrix element to the ground state.

\section{Summary}
\label{s_sum}
A consistent description
of nuclear charge radii and E0 transitions was proposed
and the validity of this approach was tested
with the interacting boson model
applied to even-even nuclei in the rare-earth region ($58\leq Z\leq74$).
An additional correlation between $\rho(E0)$ values
and the summed M1 strength to the scissors mode, valid in the SU(3) limit,
was pointed out and tested in the rare-earth region.
It would be of interest to derive
a more general relation between these two properties,
applicable to transitional and deformed nuclei.

\begin{theacknowledgments}
The work reported in this contribution
is manifestly inspired by the seminal ideas of
Akito Arima and Francesco Iachello.
I wish, on the occasion of his 70th birthday, to
express my sincere thanks to Francesco Iachello
for the encouragement and support he has given me over many years.
Conversations and exchanges with Joe Ginocchio, Achim Richter, and Peter von Neumann-Cosel
concerning the scissors M1 strength are gratefully acknowledged.
Thanks are due to Salima Zerguine and Abdelhamid Bouldjedri
in collaboration with whom part of this work was done.
Finally, I thank Kris Heyde and John Wood for stimulating discussions.
\end{theacknowledgments}

\end{document}